\pdfoutput=1
\documentclass[cernpreprint, atlasdraft=false, USenglish, coverpage=false, texmf, orcidlogo]{atlasdoc}
\usepackage{atlaspackage}
\usepackage{subfig}
\usepackage{atlasbiblatex}

\usepackage{atlasphysics}
\usepackage{tikz-feynman}
\usepackage{amsmath}
\usepackage{ANA-EXOT-2021-35-PAPER-defs}

\graphicspath{{logos/}{figures/}}

\AtlasTitle{Search for pair-produced vectorlike quarks coupling to light quarks in the lepton plus jets final state using 13 TeV $pp$ collisions with the ATLAS detector }

\AtlasAbstract{%

A search is presented for the pair-production of heavy vectorlike quarks (VLQs) that each decay into a $W$ boson and a light quark.
This study focuses on events where one $W$ boson decays into leptons and the other into hadrons.
The search analyzed 140 fb$^{-1}$ of $pp$ collision data with $\sqrt{s} = 13$ TeV, recorded by the ATLAS detector from 2015 to 2018 during Run~2 of the Large Hadron Collider.
The final state is characterized by a high-transverse-momentum isolated electron or muon, large missing transverse momentum, multiple small-radius jets, and a single large-radius jet identified as originating from the hadronic decay of a boosted $W$ boson.
With higher center-of-mass energy and integrated luminosity than in the Run~1 search, and improved analysis tools, this analysis excludes VLQs ($Q$) with masses below 1530 GeV at 95\% CL for the branching ratio $\mathcal{B}(Q\to Wq) = 1$, an improvement of 840 GeV on the previous ATLAS limit.
}

\AtlasRefCode{EXOT-2021-35}

\PreprintIdNumber{CERN-EP-2024-130}

\AtlasJournal{Phys.\ Rev.\ D.}
\AtlasJournalRef{Phys. Rev. D 110 (2024) 052009}
\AtlasDOI{10.1103/PhysRevD.110.052009}

\hypersetup{pdftitle={ATLAS document},pdfauthor={The ATLAS Collaboration}}

\begin{document}

\maketitle

\tableofcontents

\section{Introduction}
\label{sec:intro}

The Standard Model (SM) of particle physics has stood as a highly successful theory for many decades.
Despite its successes, the SM falls short in explaining several phenomena, including the matter-antimatter asymmetry, the fermion mass hierarchy, and dark matter.
A notable concern within the SM framework is the issue of the Higgs boson's mass divergence~\cite{Susskind:1978ms}, which arises from quantum corrections due to interactions with other particles.
Various models have emerged to address these shortcomings of the SM, such as Top Color~\cite{Dobrescu:1998,Hill:1995}, Little Higgs~\cite{ArkaniHamed:2002qy}, Composite Higgs~\cite{Kaplan:1983sm, Agashe:2004rs}, and the Left-Right Mirror Model ~\cite{Chakdar:2013}. A common thread among these models is the introduction of additional particles known as vectorlike quarks (VLQs), for which the left- and right-handed chiralities transform identically under the electroweak gauge group $\text{SU(2)} \times \text{U(1)}$~\cite{delAguila:1982fs, AguilarSaavedra:2009es}. These particles may either be composite states linked to the strongly coupled sector or interact with same-charge SM quarks.
At the Large Hadron Collider (LHC), a single VLQ can be produced via the electroweak interaction, whereas pair production takes place via the strong interaction. The cross-section for pair production depends only on the VLQ mass, while for single production the cross-section depends on the coupling parameter~\cite{Atre:2011ae, Aguilar-Saavedra:2013qpa, Buchkremer:2013bha}.
Most past searches have focused on models with VLQs that mix with third-generation quarks, while only a few have considered mixing with lighter generations~\cite{EXOT-2011-28, Okada:2012gy, EXOT-2014-10, CMS-B2G-12-016}. The previous best lower limit on the mass of new heavy quarks ($Q$) that decay to light SM quarks is from the CMS Collaboration, using \RunOne data~\cite{CMS-B2G-12-016}. For pair production, VLQs with masses below 845~\GeV are excluded for branching ratio $\mathcal{B}(\ensuremath{Q}\to Wq) = 1$, and for single production, masses below 685~\GeV are excluded for a weak-isospin singlet $\ensuremath{Q}$, corresponding to $\mathcal{B}(\ensuremath{Q}\to Wq:Zq:Hq) = 0.5:0.25:0.25$~\cite{Aguilar-Saavedra:2013qpa}. The previous ATLAS searches for single and pair production of such VLQs were performed at $\sqrt{s}=7$ and 8~\TeV{}~\cite{EXOT-2011-09, EXOT-2011-28, EXOT-2014-10}. In addition, the analysis in Ref.~\cite{Buckley:2020wzk} sets even stronger limits by reinterpreting and combining various channels with current LHC  measurements to VLQs.

\begin{figure}[tb]
\centering
\includegraphics[scale=0.26]{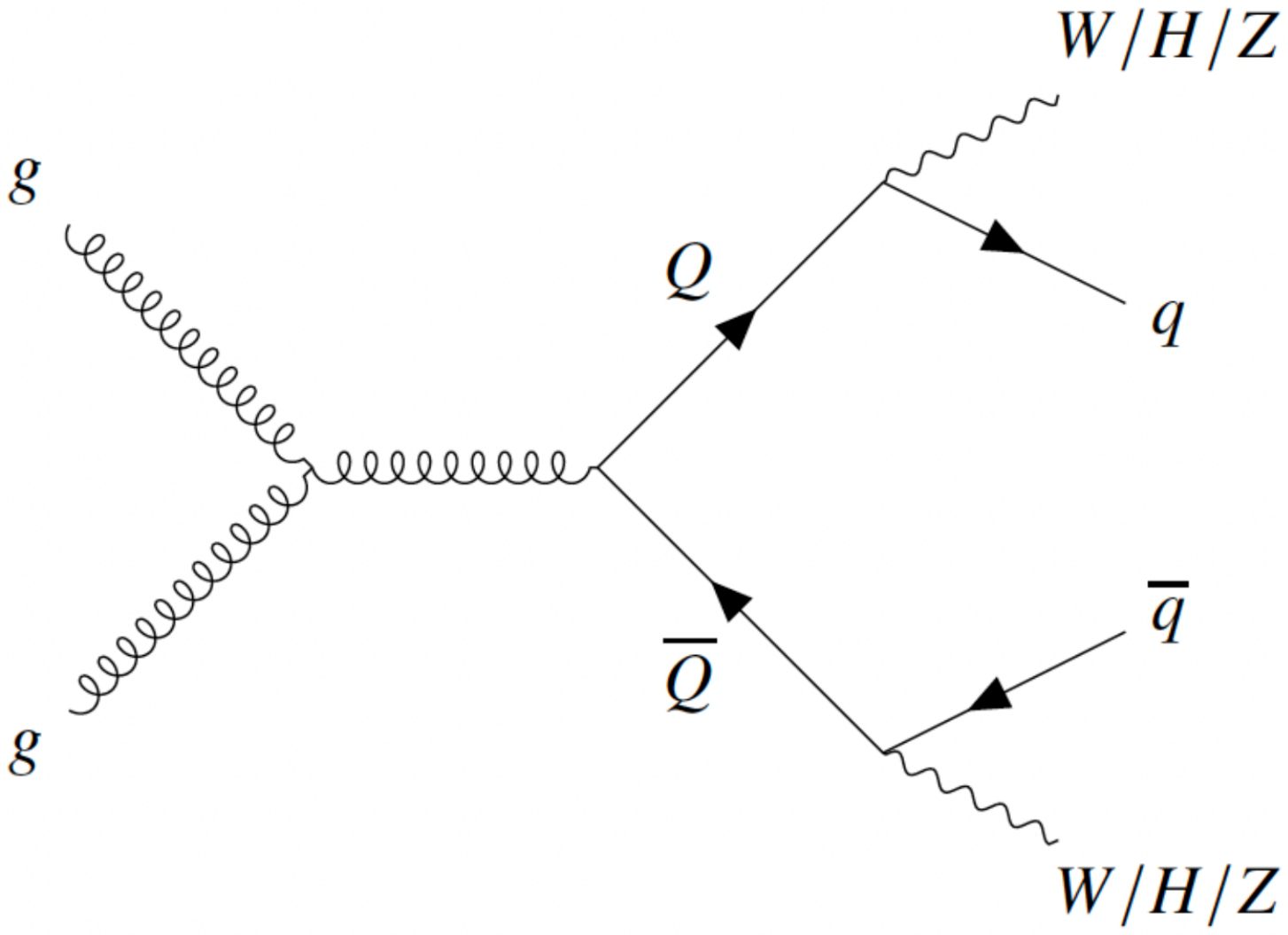}
\caption{Illustrative Feynman diagram of pair production of the vectorlike quarks. Each vectorlike quark decays into a light quark and either a $W$ boson, a $Z$ boson, or a Higgs boson.}
\label{fig:vlqFen}
\end{figure}

The present search focuses on pair production of VLQs that decay into a $W$ boson and a light quark ($\QQbar \to WqWq$, with $q = u$, $d$, or $s$) using \lumi{} of $pp$ collisions with $\sqrt{s}=13$~\TeV{} collected by the ATLAS detector.
Figure~\ref{fig:vlqFen} illustrates a characteristic Feynman diagram for this process.
Consequently, this analysis complements the more common searches for top- and bottom-partner VLQs that are assumed to mix only with the third-generation SM quarks~\cite{
EXOT-2018-58, %
EXOT-2019-08, %
WbXPLB,
CMS-B2G-17-012, %
CMS-B2G-20-011}. %
The event selection targets the final state where
one $W$ boson decays leptonically, which greatly reduces the
multijet background, and the other $W$ boson decays hadronically, which has
a large branching ratio.
Events are selected in this channel, requiring a high-$\PT{}$ charged lepton ($\ell = e$ or $\mu$), missing transverse momentum, and a minimum of three jets (comprising two small-radius jets and a large-radius jet).
For better discrimination between signal and background, the search leverages the distinct topology of the high-$\PT{}$ decay products from the VLQs, particularly the highly boosted $W$ bosons, and the transverse momentum balance between these products. The masses of VLQs where the $W$ bosons decay leptonically ($m_\text{VLQ}^\text{lep}$) and hadronically ($m_\text{VLQ}^\text{had}$) are reconstructed from the final-state particles after employing several kinematic requirements to isolate a high-purity sample of signal-like events.
The final step involves scrutinizing the number of observed events for any signal-like excess over the SM prediction.


%

%
\section{ATLAS detector}
\label{sec:detector}

%
\newcommand{\AtlasCoordFootnote}{%
ATLAS uses a right-handed coordinate system with its origin at the nominal interaction point (IP)
in the centre of the detector and the \(z\)-axis along the beam pipe.
The \(x\)-axis points from the IP to the centre of the LHC ring,
and the \(y\)-axis points upwards.
Polar coordinates \((r,\phi)\) are used in the transverse plane,
\(\phi\) being the azimuthal angle around the \(z\)-axis.
The pseudorapidity is defined in terms of the polar angle \(\theta\) as \(\eta = -\ln \tan(\theta/2)\) and is equal to the rapidity
$ y = \frac{1}{2} \ln \left( \frac{E + p_z c}{E - p_z c} \right) $ in the relativistic limit.
Angular distance is measured in units of \(\Delta R \equiv \sqrt{(\Delta y)^{2} + (\Delta\phi)^{2}}\).}

The ATLAS experiment~\cite{PERF-2007-01} at the LHC is a multipurpose particle detector
with a forward--backward symmetric cylindrical geometry and a near \(4\pi\) coverage in
solid angle.\footnote{\AtlasCoordFootnote}
It consists of an inner tracking detector (ID) surrounded by a thin superconducting solenoid
providing a \qty{2}{\tesla} axial magnetic field, electromagnetic and hadronic calorimeters, and a muon spectrometer.
The inner tracking detector covers the pseudorapidity range \(|\eta| < 2.5\).
It consists of silicon pixel, silicon microstrip, and transition radiation tracking detectors.
Lead/liquid-argon (LAr) sampling calorimeters provide electromagnetic (EM) energy measurements
with high granularity within the region \(|\eta|< 3.2\).
A steel/scintillator-tile hadronic calorimeter covers the central pseudorapidity range (\(|\eta| < 1.7\)).
The endcap and forward regions are instrumented with LAr calorimeters
for EM and hadronic energy measurements up to \(|\eta| = 4.9\).
The muon spectrometer surrounds the calorimeters and is based on
three large superconducting air-core toroidal magnets with eight coils each.
The field integral of the toroids ranges between \num{2.0} and \qty{6.0}{\tesla\metre}
across most of the detector.
The muon spectrometer includes a system of precision tracking chambers up to \(|\eta| = 2.7\) and fast detectors for triggering up to \(|\eta| = 2.4\).
The luminosity is measured mainly by the LUCID--2~\cite{LUCID2} detector, which is located close to the beampipe.
A two-level trigger system is used to select events~\cite{TRIG-2016-01}.
The first-level trigger is implemented in hardware and uses a subset of the detector information
to accept events at a rate below \qty{100}{\kHz}.
This is followed by a software-based trigger that
reduces the accepted event rate to \qty{1}{\kHz} on average
depending on the data-taking conditions.
A software suite~\cite{SOFT-2022-02} is used in data simulation, in the reconstruction
and analysis of real and simulated data, in detector operations, and in the trigger and data acquisition
systems of the experiment.


%
%
\section{Data and simulated event samples}
\label{sec:samples}

The analysis uses the ATLAS \RunTwo data collected during 2015--2018, with an integrated luminosity of $140.1\pm 1.2~\mathrm{fb}^{-1}$~\cite{DAPR-2021-01,LUCID2} from
$pp$ collisions at a center-of-mass energy of $\sqrt{s}=13$~\TeV{}.
Data quality requirements~\cite{DAPR-2018-01} were applied while choosing the events to be analyzed. These requirements include stable-beam conditions with all detector subsystems and relevant components operational while recording the data.

Monte Carlo (MC) simulated event samples are used to model the background and signal events. All samples were produced using the ATLAS simulation infrastructure~\cite{SOFT-2010-01} and \GEANT~\cite{Agostinelli:2002hh} to simulate the response of the detector.  Data-driven corrections were applied to simulated \Wjets{}, $t\bar{t}$, single-top and multijet events, while all other processes were estimated in a purely MC-driven way.
The effects of \pileup, which refers to additional proton-proton interactions in an event, were modeled by overlaying the hard-scattering events with minimum-bias events, simulated using the soft QCD processes of \PYTHIA[8.186]~\cite{Sjostrand:2014zea} with the A3 set of tuned parameters (\enquote{tune})~\cite{ATL-PHYS-PUB-2016-017}.
Differences between the \pileup conditions in data and simulation are taken into account by reweighting the mean number of interactions per bunch crossing in simulation to the one observed in data.

Signal samples simulating pair-produced vectorlike quarks $\QQbar{}$ were generated at leading order (LO) with \MADGRAPH[5]~\cite{Alwall:2011uj} using the \NNPDF[3.0nlo]~\cite{Ball:2014uwa} parton distribution function (PDF) set, and interfaced with \PYTHIA[8.186] to model the parton shower, hadronization and underlying event.
The samples were produced for the weak-isospin singlet model in the narrow-width approximation with masses from 800~\GeV to 2~\TeV, in steps of 100~\GeV.
Dedicated signal samples were also produced for a VLQ from a weak-isospin doublet for the 1.2~\TeV mass point and used to confirm that kinematic differences between the singlet and doublet models have a negligible effect on the results.
The final results are also tested for other branching ratio scenarios by performing event-by-event reweighting using the generator's decay information.
The signal sample cross-sections were calculated with \TOPpp[2.0]~\cite{Czakon:2011xx} at next-to-next-to-leading order (NNLO) in QCD including the resummation of next-to-next-to-leading logarithmic soft-gluon terms.

The production of a single $W$ boson in association with jets is the dominant background.
The production of $V+$\,jets ($V=W,Z$) was simulated with the
\SHERPA[2.2.11]~\cite{Bothmann:2019yzt}
generator using next-to-leading-order (NLO) matrix elements (ME) for up to two partons, and LO matrix elements
for up to five partons, calculated with the Comix~\cite{Gleisberg:2008fv}
and \OPENLOOPS~\cite{Buccioni:2019sur,Cascioli:2011va,Denner:2016kdg} libraries. They
were matched with the \SHERPA parton shower~\cite{Schumann:2007mg} using the \MEPSatNLO
prescription~\cite{Hoeche:2011fd,Hoeche:2012yf,Catani:2001cc,Hoeche:2009rj} and the set of tuned parameters developed by the \SHERPA authors.
The \NNPDF[3.0nnlo]~\cite{Ball:2014uwa} PDF set was used and the samples
were normalized to a NNLO prediction~\cite{Anastasiou:2003ds}. Generator-level \Wjets{} samples were also produced with the Catani-Krauss-Kuhn-Webber (CKKW)  merging/matching scale shifted from 20~\GeV to 15~\GeV or 30~\GeV, or with the resummation scale (QSF) parameters varied from 1 to 0.25 or 4, to estimate the related uncertainties.

The production of \ttbar events was modeled using the
\POWHEGBOX[v2]~\cite{Frixione:2007nw,Nason:2004rx,Frixione:2007vw,Alioli:2010xd}
generator at NLO with the \NNPDF[3.0nlo] PDF set
and the \hdamp parameter\footnote{The
\hdamp parameter is a resummation damping factor and one of the
parameters that control the matching of \POWHEG matrix elements to
the parton shower and thus effectively regulates the
high-\pT radiation against which the \ttbar system recoils.} set
to 1.5\,$m_{\text{top}}$~\cite{ATL-PHYS-PUB-2016-020}.  The events were interfaced
to \PYTHIA[8.230] to model the parton shower,
hadronization, and underlying event, with parameter values set according
to the A14 tune~\cite{ATL-PHYS-PUB-2014-021} and using the \NNPDF[2.3lo]
set of PDFs~\cite{Ball:2012cx}.
The impact of using a different parton shower and hadronization model was evaluated by replacing \PYTHIA[8.230] with \HERWIG[7.0]~\cite{Bahr:2008pv,Bellm:2015jjp},
using the H7UE tune~\cite{Bellm:2015jjp} and the
\MMHT[lo] PDF set~\cite{Harland-Lang:2014zoa}.
To assess the uncertainty in the matching of NLO matrix elements to the
parton shower, the \POWHEG sample was compared with a sample of events
generated with \MGNLO[2.6.0]~\cite{Alwall:2014hca} interfaced with
\PYTHIA[8.230]. The \MGNLO calculation used the
\NNPDF[3.0nlo] PDF set, and \PYTHIA[8] used the \NNPDF[2.3lo] PDF set
with the A14 tune.
The potential impact of underestimating the amount of initial-state radiation (ISR) is assessed by comparing the nominal sample with another that increases the \hdamp value to 3\,$m_{\text{top}}$, halves the renormalization and factorization scales, and utilizes the \enquote{Var3cUp} weight from the A14 tune. Conversely, a possible overestimation of the amount of ISR is examined by doubling the renormalization and factorization scales in another sample and selecting the \enquote{Var3cDown} weight from the A14 tune. Additionally, the uncertainty related to modeling of final-state radiation (FSR) is evaluated by either doubling or halving the renormalization scale for emissions from the parton shower.

Background events may also originate from the production of single-top-quark events. The main contribution to this background comes from top quarks associated with a $W$ boson ($tW$), while $t$- and $s$-channel single-top production gives only a minor contribution.
The single-top background was modeled by the \POWHEGBOX[v2]~\cite{Re:2010bp,Nason:2004rx,Frixione:2007vw,Alioli:2010xd} generator at NLO in QCD using the five-flavor scheme and the
\NNPDF[3.0nlo] set of PDFs. The events were interfaced with \PYTHIA[8.230], which used the A14 tune.
Comparisons with samples generated by \MGNLO[2.6.2] at NLO in QCD are used to estimate the uncertainty in the matching of NLO matrix elements to the parton shower. The parton shower and hadronization uncertainties are evaluated from comparisons with samples generated with an alternative showering program, \HERWIG[7.04], using the H7UE tune and the \MMHT[lo] PDF set.
For the $tW$ samples, the diagram removal (DR) scheme~\cite{Frixione:2008yi} was used to remove interference and overlap with \ttbar production.
The related uncertainty was estimated by comparing these samples with an alternative sample
generated using the diagram subtraction (DS) scheme~\cite{Frixione:2008yi,ATL-PHYS-PUB-2016-020}.

The contributions from other processes are nearly negligible relative to the overall background.
The largest additional background contribution comes from the production of two bosons,
either two vector bosons ($VV$) or a vector boson and a Higgs boson ($VH$), with at least one lepton in the final state.
Diboson events with multiple leptons in the final state
are strongly disfavored by requiring exactly one reconstructed lepton in the final state.
The $VV$ samples were generated with \SHERPA[2.2.2], whereas the $VH$ samples were generated with \PYTHIA[8.186].
Fully leptonic final states and semileptonic final states, where one boson
decays leptonically and the other hadronically, were generated using
matrix elements at NLO accuracy in QCD for up to one additional parton
and at LO accuracy for up to three additional parton
emissions. Samples for the loop-induced $gg \to VV$ processes were
generated using LO-accurate matrix elements for up to one
additional parton emission for both the fully leptonic and
semileptonic final states. The \MEPSatNLO{} prescription was used to
match and merge the matrix element calculations with the \SHERPA parton shower based on Catani--Seymour
dipole factorization~\cite{Gleisberg:2008fv,Schumann:2007mg}.
The virtual QCD corrections were provided by the
\OPENLOOPS library. The
\NNPDF[3.0nnlo] set of PDFs was used, along with a dedicated set of tuned parton-shower parameters developed by the
\SHERPA authors.
Finally, production of $t\bar{t}V$ events was modeled using the
\MGNLO[2.3.3] generator at NLO with the
\NNPDF[3.0nlo] PDFs, interfaced to \PYTHIA[8.210] using the A14 tune and the
\NNPDF[2.3lo] PDF set.
Multijet events were simulated using the \SHERPA[2.1.1] generator with the default \CT[10] PDF set~\cite{Lai:2010vv}.


%

%
%

%
\section{Object reconstruction}
\label{sec:object}

%
Electrons are reconstructed from energy clusters in the EM calorimeter matched with ID tracks.
Electron candidates must be located in the central region of the detector ($\left| \eta \right| < 2.47$), have $\pt{} > 27$~\GeV, and be matched to a track with $\left| z_0 \sin \theta \right| < 0.5$~mm and $\left| d_0 / \sigma_{d_0} \right| < 5$, where $d_0$ is the track's transverse impact parameter with respect to the hard-scatter vertex and its uncertainty is given by $\sigma_{d_0}$, and $z_0$ is its longitudinal impact parameter.
Candidates in the transition region between the EM calorimeter's barrel and endcap sections ($1.37 < \left| \eta \right| < 1.52$) are excluded.
A "baseline electron" selection requires electrons to satisfy the \textit{Medium} likelihood identification criteria with no selection on the isolation. Alternatively, a "tight electron" selection requires satisfying the \textit{Tight} likelihood identification criteria, with loose isolation requirements~\cite{EGAM-2018-01}.
The first isolation requirement is $E_{\text{T, cone}}^{\text{isol}} / \pt{e} < 0.2$, where $\pt{e}$ is the electron candidate's \pt{}, and $E_{\text{T,cone}}^{\text{isol}}$ is the energy deposited in the calorimeter within a cone of angular size $\Delta R = 0.2$ around the candidate's direction, with energy leakage and \pileup contributions subtracted.
The second requirement is $p_{\text{T,var}}^{\text{isol}} / \pt{e} < 0.15$, where  $p_{\text{T,var}}^{\text{isol}}$ is the sum of the track \pt{}, excluding the electron candidate, within a cone of size $\Delta R=\min(0.2, \SI{10}{GeV} / \pt{e})$.
Data-based scale factors~\cite{EGAM-2018-01} are employed to correct for variations in reconstruction, identification, isolation, and trigger efficiencies between data and simulation.

Muons are reconstructed~\cite{MUON-2018-03} using combined tracks in the MS and the ID. Baseline muons must meet \textit{Loose} identification criteria with no isolation requirements, while tight muons must satisfy \textit{Tight} identification criteria~\cite{MUON-2018-03} and meet the track-based isolation requirements defined by the \enquote{TightTrackOnly} working point.
This working point uses the scalar sum of the \pt{} of all tracks that are within a cone of size $\Delta R = \min(0.3, \SI{10}{\GeV}/\pt{\mu})$ around the muon candidate, where $\pt{\mu}$ is the muon candidate's \pt{}.
The track matched to the muon candidate under consideration is excluded from the sum.
The muon is selected if this sum is less than 6\% of $\pt{\mu}$. All muon candidates must also satisfy $|z_0 \sin\theta| < 0.5$~mm and  $\left| d_0 / \sigma_{d_0} \right| < 3$.
Muons are required to have $\pt{}> 25$~\GeV, and a reconstruction limit of $|\eta|<2.5$.
Efficiency scale factors are applied to account for differences in muon reconstruction, identification, vertex association, isolation, and trigger efficiencies between simulation and data~\cite{MUON-2018-03}.

Small-radius (small-$R$) jet candidates are formed from particle-flow objects~\cite{PERF-2015-09}, utilizing the \antikt algorithm~\cite{Cacciari:2008gp,Fastjet} with a radius parameter of $R=0.4$.
The particle-flow algorithm combines information from ID tracks and calorimeter energy deposits to construct input for jet reconstruction.
The jet energy is calibrated to the particle scale, i.e.\ without detector effects, through a series of corrections, including simulation-based adjustments and in situ calibrations~\cite{JETM-2018-05}.
Jets are required to have $\pt{} > 25$~\GeV and $|\eta| < 2.5$.
To reject jets originating from pile-up interactions, jet candidates with $|\eta| < 2.4$ and $20 < \pt{} < 60$~\GeV must satisfy the tight jet-vertex-tagger (JVT) criterion~\cite{PERF-2014-03}.
Small-$R$ jets containing a $b$-hadron decay are identified using the DL1r algorithm~\cite{ATLAS:2022qxm}. Jets are considered $b$-tagged if they meet the operating point criteria for 70\% efficiency~\cite{FTAG-2018-01}.
The $b$-tagging efficiencies in simulation, as well as the charm and light mistag rates, are corrected to match those in data~\cite{FTAG-2020-08, arXiv:2301.06319}.

Large-radius (large-$R$) jets are constructed from noise-suppressed topological calorimeter-cell clusters with the \antikt{} algorithm with a radius parameter of $R=1.0$, and calibrated using local hadronic cell reweighting~\cite{PERF-2014-07}.  These \largeRjet{s} are required to have $\pt{} > 200$~\GeV and $|\eta| < 2$.

A $W$-boson tagging algorithm identifies high-$\pt{}$ hadronically decaying $W$ bosons that produce a single collimated \largeRjet~\cite{JETM-2018-03}, with 80\% efficiency. It uses criteria based on the mass of the \largeRjet, the number of ID tracks associated with the jet, and the energy correlation function ratio $D_{2}$~\cite{JETM-2018-03,Larkoski_2014}. Scale factors adjust the $W$-boson tagging efficiency in simulation to match that in data~\cite{ATL-PHYS-PUB-2020-017}.

The missing transverse momentum is calculated as the negative vectorial sum of the transverse momenta of all the calibrated objects in an event, augmented by a track-based soft-term that accounts for energy depositions linked to the event's primary vertex but not attributed to any calibrated object~\cite{JETM-2020-03}. The magnitude of the missing transverse momentum is denoted by $\met$ and is required to exceed 250~\GeV. In signal events, a significant amount of \met is expected from the boosted leptonically decaying $W$ boson.

An overlap removal procedure, based on the baseline lepton definitions, is employed to avoid double-counting of ambiguous reconstructed objects.
First, electron--muon overlap is addressed by removing any muon that shares a track in the ID with an electron if the muon is only \enquote{calorimeter-tagged} (because of poor MS acceptance at $|\eta|\approx 0$), or otherwise removing the electron.
Subsequently, overlap between jets and leptons is resolved by rejecting any jets within $\Delta R = 0.2$ of an electron, and then rejecting any electrons within $\Delta R = 0.4$ of a jet.
Similarly, jets are rejected if they have fewer than three associated tracks and are within $\Delta R = 0.2$ of a muon candidate; otherwise, the muon is rejected if it lies within $\Delta R = \text{min}(0.4, 0.004 + \SI{10}{\GeV}/\pt{\mu})$ of a jet.

An additional procedure to resolve overlaps between \smallRjet{s} and \largeRjet{s} is described in Section~\ref{sec:event}.


%
%
\section{Event selection}
\label{sec:event}

This analysis focuses on the VLQ pair decay $\QQbar{} \to WqWq$ with one $W$ boson decaying leptonically and the other hadronically.
A single-lepton trigger~\cite{TRIG-2018-05, TRIG-2018-01} was used and events selected with this trigger were required to have exactly one electron with $\pT{} \geq 27~\GeV$ or one muon with $\pT{} \geq 25~\GeV$. In addition to this, at least three \smallRjet{s} with $\pT{} \geq 25~\GeV$, and $\met \geq 250~\GeV$, are required. These basic event-level requirements ensure the identification of at least one quark from a hadronically decaying $W$ boson, two additional quarks coming directly from the two VLQ decays, and one leptonically decaying $W$ boson.

Events must satisfy a series of criteria referred to as the \enquote{preselection} that significantly reduce the background while retaining events consistent with the signal process.
As summarized in Table~\ref{tab:preselection}, the preselection requires exactly one charged lepton, large \ETmiss{}, at least one \largeRjet, and at least two \smallRjet{s}.
Events passing preselection have two reconstructed $W$ candidates.
The leptonically decaying $W$ boson (\Wlep) is reconstructed by combining the lepton and the reconstructed neutrino; where the neutrino is reconstructed using the event's $\MET$ and the $z$-component of the neutrino's momentum, estimated by using the mass of the $W$ boson as a constraint.
The hadronically decaying $W$ boson (\Whad), is defined as the leading $W$-tagged \largeRjet.  If no \largeRjet{} meets the $W$-tagging criteria, then the \Whad{} candidate is
defined to be the \largeRjet{} closest in mass to the $W$ boson~\cite{ParticleDataGroup:2022pth}.
Finally, an additional overlap removal procedure requires \smallRjet{s} to be separated from the \Whad{} candidate by $\Delta R \geq 1.0$. Small-$R$ jets  that fail this requirement are removed from the analysis and thus not considered in the reconstruction of the VLQs.

The input for the VLQ pair reconstruction algorithm consists of the \Whad{} and \Wlep, and the \smallRjet{s} after overlap removal. The three leading \smallRjet{s} are then paired with the reconstructed $W$ bosons, and for each pairing the invariant mass of each VLQ candidate is computed. The pairing resulting in the smallest mass difference \dm{} is chosen.

Events that pass the preselection are then categorized into signal regions (SRs), reweighting regions (RwRs), and validation regions (VRs).
These regions are orthogonal, and the RwRs and VRs are chosen to be kinematically close to the SRs.
The SRs are designed to maximize the sensitivity to the possible presence of signal events.
The RwRs have high purity for a particular SM process and are used to calculate data-driven corrections for that process.
The VRs are also dominated by a single SM process and are used to validate the MC simulation's modeling of the data after corrections.
Kinematic requirements for the variables defining the SRs are chosen to optimize $S/\sqrt{B}$ for a signal mass of 1400~\GeV, since this is close to the expected mass reach of the analysis for signal events in the $WqWq$ decay channel.

Two signal regions, SR1 and SR2, are defined for this analysis. Using two SRs increases the sensitivity since SR1 has a higher $S/\sqrt{B}$ than when using a single SR combining SR1 and SR2.
Events in either SR must have zero $b$-tagged \smallRjet{s}, at least one $W$-tagged \largeRjet, $\Delta\phi(\text{lepton}, \met) \leq 0.5$, $\Delta R (\Wlep, \Whad) \geq 0.8$ and $\ST \geq 2000$~\GeV, where \ST{} is the scalar sum of the \met and the transverse momenta of the charged lepton and selected \smallRjet{s}. This variable has higher discriminating power due to the large expected mass of the VLQ.
A final requirement separates the events into either SR1 or SR2. SR1 requires $\Delta\phi(\text{lead jet}, \met) < 2.75$, resulting in a higher $S/\sqrt{B}$. SR2 requires $\Delta\phi(\text{lead jet}, \met) \ge 2.75$, selecting signal events not assigned to SR1, at the expense of a small contribution from multijet events.

Three reweighting regions, \WRwR{}, \ttRwR{}, and multijetRwR, are used to correct for mismodeling in the \Wjets~\cite{ATL-PHYS-PUB-2017-006}, \ttbar{} plus single-top, and multijet background predictions, respectively.
In addition, the validation regions \WVR{}, \ttVR{}, and multijetVR are defined in order to validate the modeling of the \Wjets, \ttbar{}, and multijet backgrounds in independent regions.
Kinematic requirements for the SRs, RwRs and VRs are summarized in Table~\ref{tab:regions} in addition to the usage of region in the fit model of the statistical analysis.
The normalization of the multijet MC samples is determined using multijetRwR.
Both \ttRwR{} and \WRwR{} are used to correct the normalization and shape of the respective MC event distributions.

\begin{table}[tb]
\centering
\caption{Common preselection criteria for all analysis regions.}
\label{tab:preselection}
\footnotesize
\begin{tabular}{c}
\toprule
=1 isolated lepton ($\ell = e$ or $\mu$) with $\pt{\ell} \ge 60~\GeV{}$ \\
$\MET\geq 250$~\GeV{} \\
$\ge 1$ \largeRjet{} with $\pt{} \ge 200~\GeV{}$ \\
$\ge 2$ \smallRjet{}\textnormal{s} with $\pt{} \ge 25~\GeV{}$ \\
$\ge 1$ \smallRjet{} with $\pt{} \ge 200~\GeV{}$ \\
$\Delta R(\smallRjet{}\textnormal{s},\Whad{})>1.0$
\\
\bottomrule
\end{tabular}
\end{table}

\begin{table}[tb]
\centering
\caption{Selection criteria for the two signal regions and three reweighting and validation regions used in the analysis.}
\label{tab:regions}
\footnotesize
\begin{tabular}{lccccccc}
\toprule
Variable & SR1 (SR2) & multijetRwR & multijetVR & \WRwR{} & \WVR{} & \ttRwR{} & \ttVR{} \\
\midrule
$N_{b\text{-tags}}$ & $=0$~~\, & $=0$~~~ & $=0$~~\, & $=0$~~~ & $=0$~~\, & $\geq 1$~~~ & $\geq 1$~~~ \\
$N_{W\text{-tags}}$ & $\geq 1$~~~ & $=0$~~~ & $\geq 1$~~~ & $=0$~~~ & $> 0$~~~ & $\geq 1$~~~ & $=0$~~~ \\
$\Delta R(W_{\text{lep}}, W_{\text{had}})$ & $\geq 0.8$ & $< 0.8$ & $< 0.8$ & $\geq 0.8$ & $< 0.8$ & $< 0.8$ & $\geq 0.8$ \\
$\Delta \phi(\text{lepton}, \MET)$ & $\leq 0.5$ & $\leq 0.1$ & $\leq 0.1$ &  & $> 0.1$ & & \\
$\ST$ & $\geq 2000$~\GeV~ & & & & & & \\
$\Delta \phi(\text{leading jet}, \MET)$ & $< 2.75$ ($\geq 2.75$) & & & & & & \\
\midrule
Included in Fit & Yes & No & No & No & No & No & No \\
\bottomrule
\end{tabular}
\end{table}


%
%
\section{Background estimation}
\label{sec:bkg}

The background for this analysis is primarily due to \Wjets{} events, followed by \ttbar{} and single-top events. Except for these and the multijet backgrounds, all other backgrounds such as \Zjets{},
$ttV$, and diboson ($VV$ and $VH$) events are combined due to their small contributions and labeled as \enquote{Other Bkgs}.  These other backgrounds are estimated purely from the MC simulations.
Utilizing the RwRs described above, reweighting factors are derived for \Wjets{} events, multijet events, and the combined \ttbar{} plus single-top events. This reweighting is performed to correct the normalization and shape of the \ST{} distribution for the \Wjets{} and \ttbar{} plus single-top events. The reweighting is determined by fitting the ratio of the observed data, after subtracting the SM predictions for processes other than the one being corrected, to the prediction for the process being corrected.
A normalization reweighting factor is derived for the multijet background, while the normalization and shape reweightings for \Wjets{} events and the combined  \ttbar{} plus single-top events pertain to their \ST{} distributions, with the above-described ratio fitted to a function of form:

\begin{equation}\label{eq:fit_func}
f(\ST{}) = P_0 + \exp(P_1 \ST{}),
\end{equation}
where $P_{0}$ and $P_{1}$ are the fitted parameters. The reweightings are derived in an iterative procedure, which terminates when all reweightings change by less than 1\% from the previous iteration.
In the first step of the first iteration, the multijet normalization reweighting is derived in the multijetRwR.
In the second step of the first iteration, the previously derived multijet reweighting is applied to multijet events and then the \ttRwR{} is used to derive the reweighting function for \ttbar{} and single-top events.
For the last step of the first iteration, the previously derived reweightings are applied to the multijet events and the \ttbar{} and single-top events, and the reweighting for the \Wjets{} background is derived in the \WRwR{} and applied to the \Wjets{} events.
These steps are repeated in the next iteration. However, at the start of each step of this iteration, only reweightings derived for samples other than the one under consideration are applied to the relevant samples. The previously derived reweighting for the sample under consideration in the step is not used.
For example, in the first step of the second iteration, the multijet normalization reweighting is rederived, but with only the reweightings obtained for \ttbar{} plus single-top and \Wjets{} from the first iteration applied to the \ttbar{}, single-top, and \Wjets{} samples.
This procedure was found to converge after just two iterations, and a third iteration produced reweighting factors almost identical to those obtained in the second iteration.

\begin{figure}[tb]
\centering
\subfloat[]{
\includegraphics[width=0.45\textwidth]{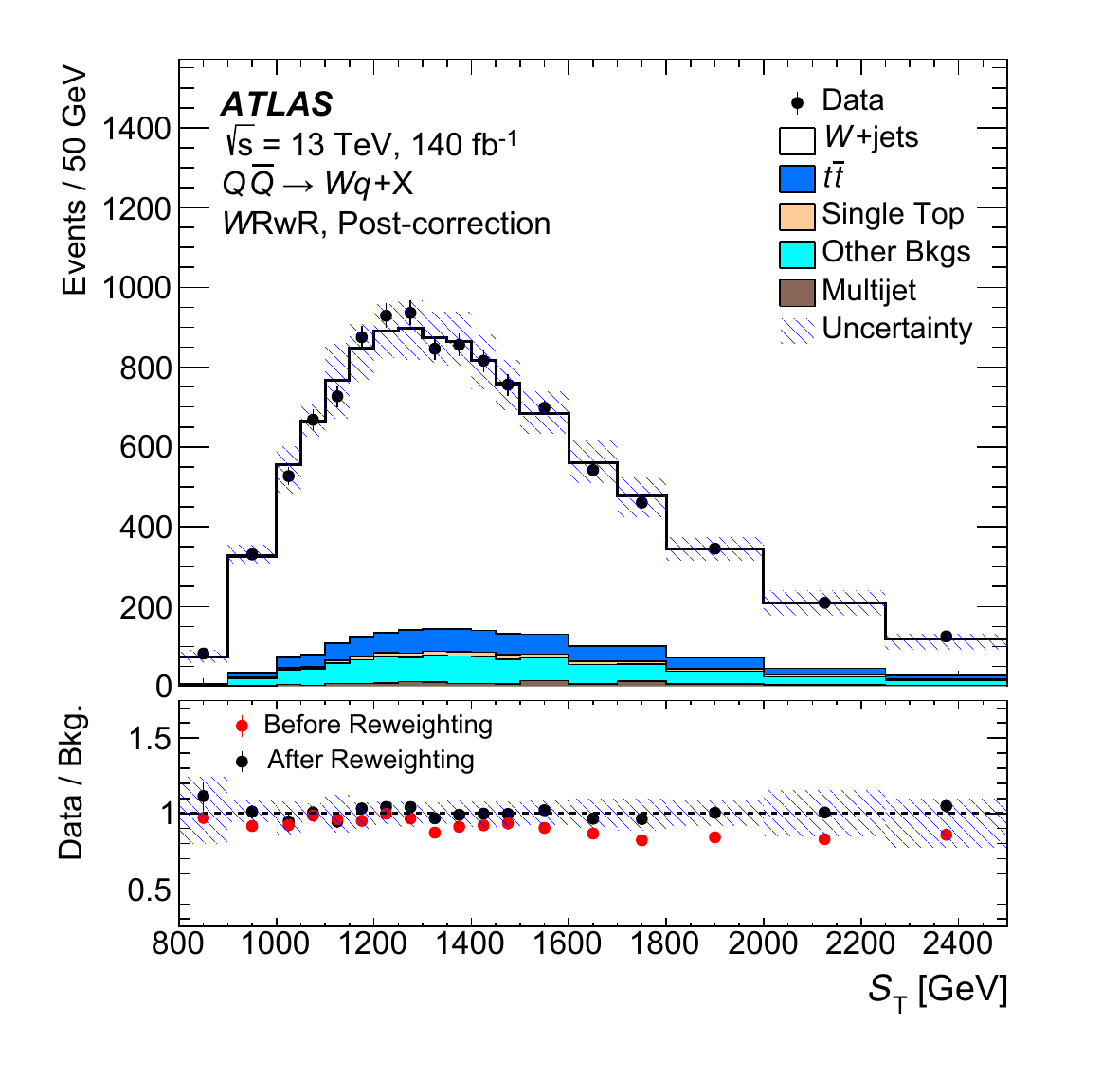}
\label{fig:corrST_WRwR}
}
\subfloat[]{
\includegraphics[width=0.45\textwidth]{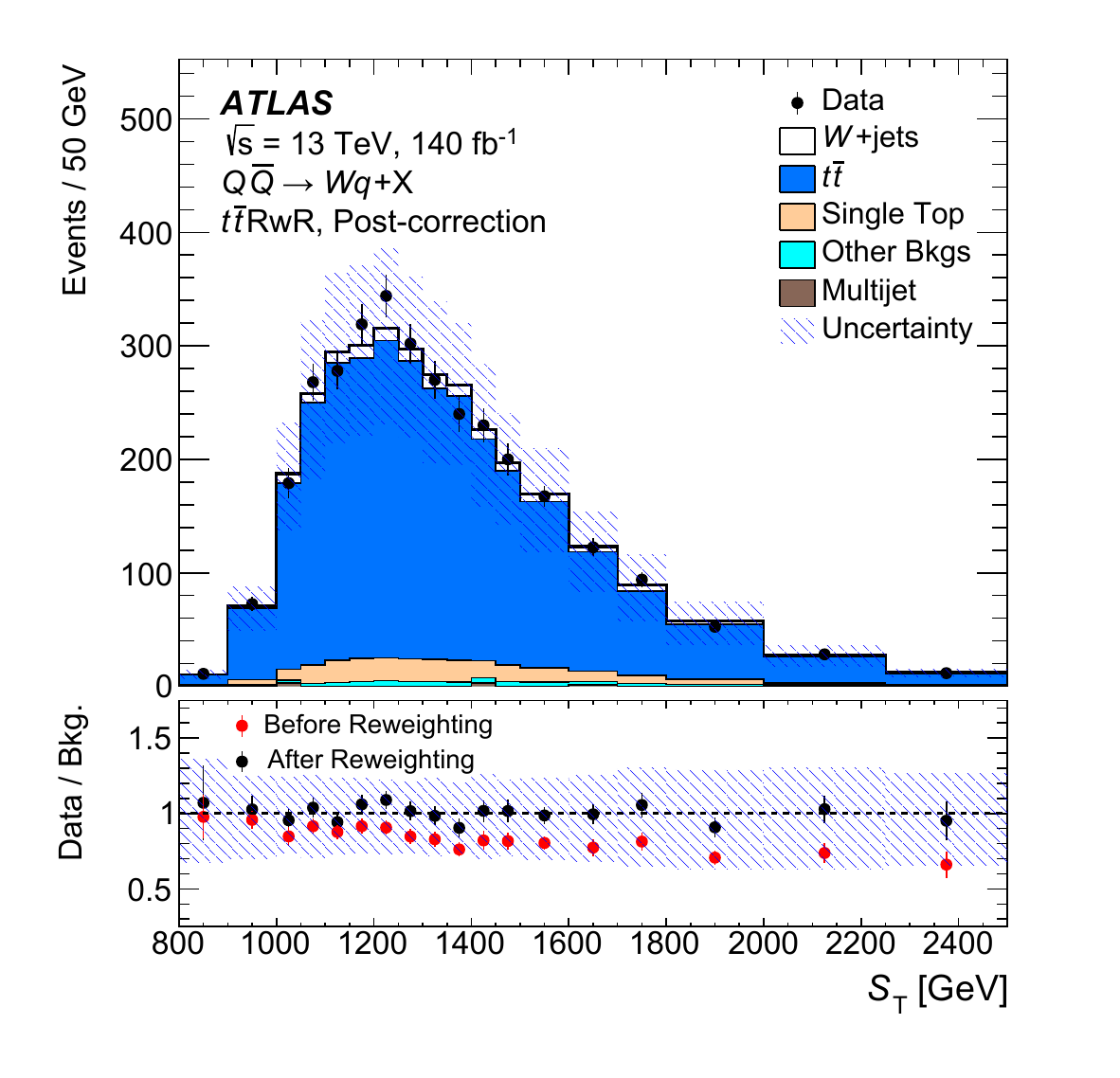}
\label{fig:corrST_ttbarRwR}
}
\caption[RwR Closure Plots]{Reconstructed \ST{} distribution  in the \protect\subref{fig:corrST_WRwR} \WRwR{} and \protect\subref{fig:corrST_ttbarRwR} \ttRwR{} after applying the reweighting corrections.  The uncertainty band includes all systematic uncertainties, as described in detail in Section~\ref{sec:systematics}. The lower panel shows the ratio of data to the SM prediction before (red points) and after (black points) applying the reweighting corrections.}
\label{fig:st_RwR}
\end{figure}

\begin{figure}[tb]
\centering
\subfloat[]{
\includegraphics[width=0.45\textwidth]{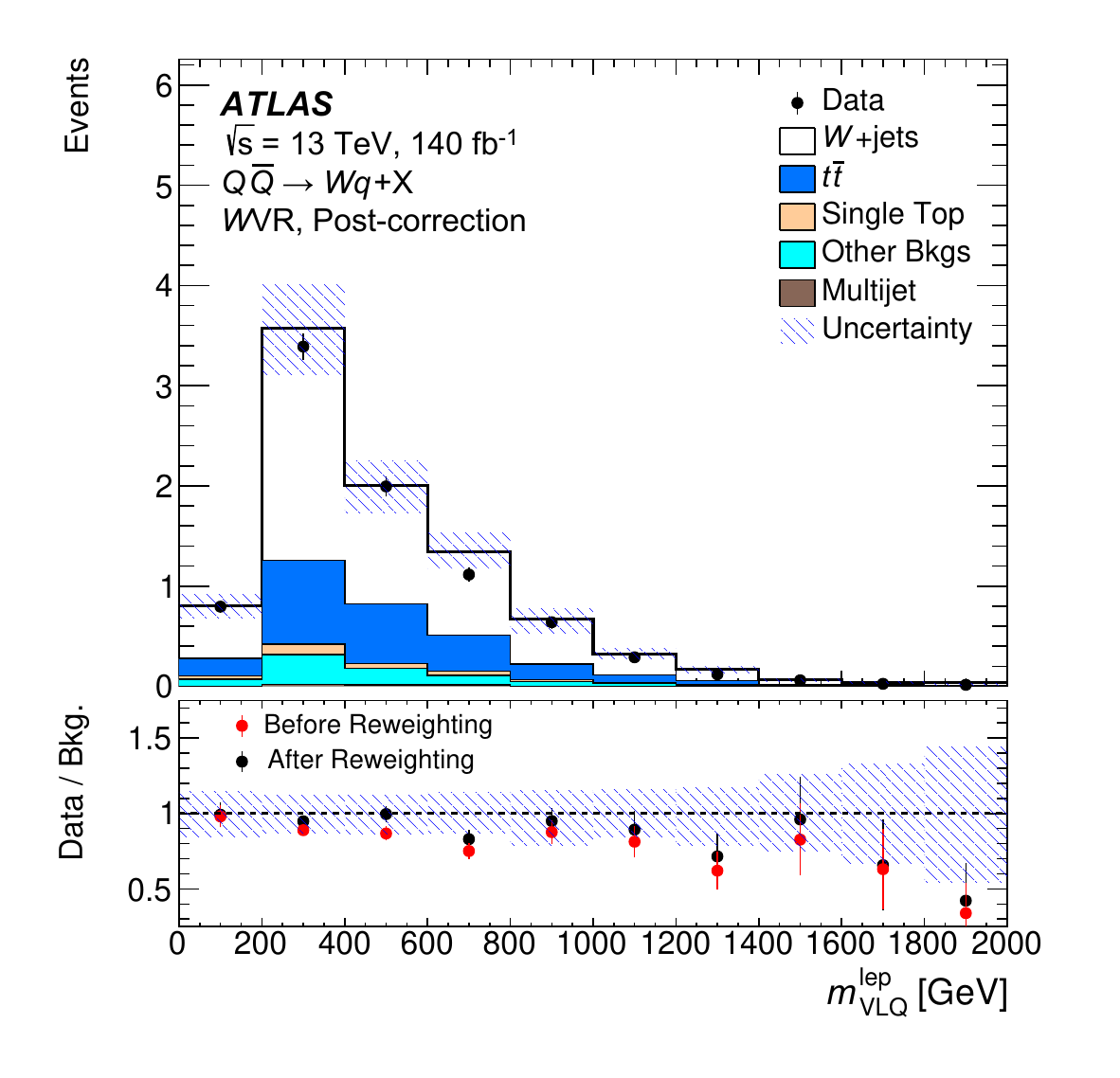}
\label{fig:mvlq_WVR}
}
\subfloat[]{
\includegraphics[width=0.45\textwidth]{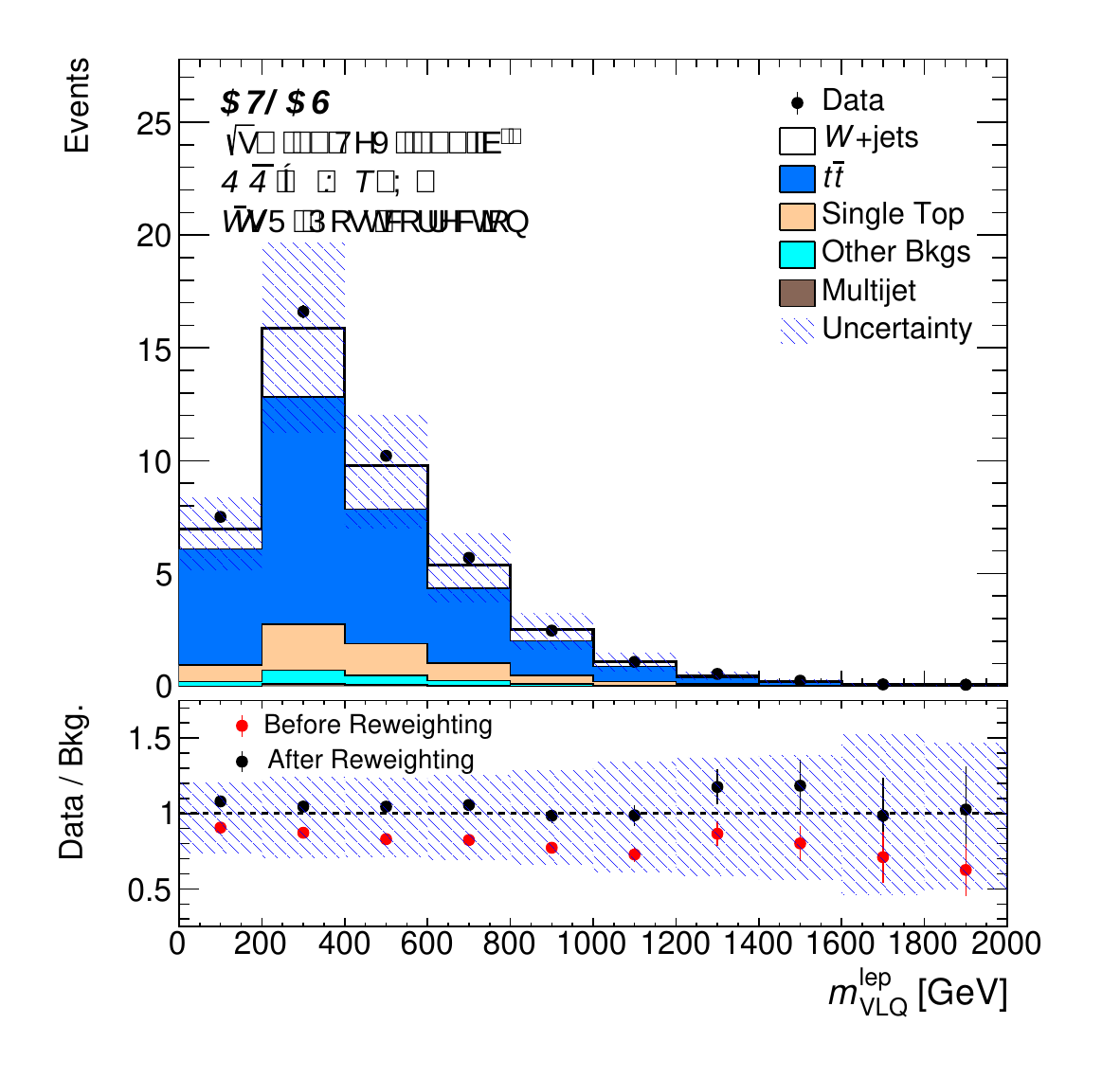}
\label{fig:mvlq_ttbarVR}
}
\caption[VR Plots]{Reconstructed mass distribution of leptonically decaying VLQ candidates in the \protect\subref{fig:mvlq_WVR} \WVR{} and \protect\subref{fig:mvlq_ttbarVR} \ttVR{} after applying the reweighting corrections.  The uncertainty band includes all systematic uncertainties, as described in detail in Section~\ref{sec:systematics}. The lower panel shows the ratio of data to the SM prediction before (red points) and after (black points) applying the reweighting corrections.}
\label{fig:mvlq_VR}
\end{figure}

Figure~\ref{fig:st_RwR} shows the \ST{} distribution in the \WRwR{} and \ttRwR{} after applying the reweighting for \Wjets{}, \ttbar{}, and multijet events as determined in their respective reweighting regions. The red marker in the data-to-background ratio represents the ratio before the reweighting is applied, whereas the black marker represents the ratio after the reweighting is applied.
The derived reweighting factors improve the modeling in the high \ST{} regions. The mass of leptonically decaying VLQ candidates is used as a discriminating variable in the final likelihood fit. Improvement in modeling is further validated in Figure~\ref{fig:mvlq_VR} by plotting this discriminating variable, \mQlep{}, in the respective VRs for \Wjets{} and \ttbar{} backgrounds after applying the reweighting, and showing the Data/MC ratios before and after the reweighting.

%
%
\section{Systematic uncertainties}
\label{sec:systematics}

Systematic uncertainties can be categorized into two main types: modeling uncertainties and experimental uncertainties.
Modeling uncertainties encompass uncertainties related to the modeling of specific physical processes and the parameters chosen within those models. These uncertainties are often related to calculations and assumptions in the underlying theory.
Experimental uncertainties concern detector setup, the modeling of the detector's response to the various objects and the uncertainties in the data-driven corrections.

The uncertainty in the combined 2015--2018 integrated luminosity is 0.83\% \cite{DAPR-2021-01}, obtained using the LUCID-2 detector~\cite{LUCID2} for the primary luminosity measurements.
\Pileup corrections are performed by adjusting the \pileup distribution in simulated data to match the real data. Uncertainties are then calculated by varying the correction.

Uncertainties related to electron and muon energy-scale calibration and resolution, and electron and muon reconstruction, identification, and isolation efficiencies are considered~\cite{EGAM-2018-01, MUON-2022-01, MUON-2018-03}. Variations are also applied to account for uncertainties in the muon spectrometer's track resolution and momentum scale.

Uncertainties in the jet energy scale and jet energy resolution~\cite{JETM-2018-05} are taken into account for both \smallRjet{s} and \largeRjet{s}. In addition, the jet mass scale~\cite{JETM-2018-02} and jet mass resolution~\cite{ATLAS-CONF-2020-022} are considered for both the small-$R$ and \largeRjet{s}. The systematic uncertainty associated with the JVT requirement is estimated by varying the scale factor used to raise or lower the JVT efficiency in simulation within its uncertainties.

The $b$-tagging efficiency corrections include uncertainties from the $b$-tagged-jet selection. An uncertainty in extrapolating of the $b$-tagging-efficiency calibration to high jet $\pt{}$ is considered. Systematic uncertainties in the $W$-tagging efficiency and inefficiency corrections are applied~\cite{JETM-2018-03}.

Uncertainties in data-driven corrections to simulated events, as described in Section~\ref{sec:bkg} for processes such as multijets and $t\bar{t}$, are considered in the analysis.
For \ttbar{} and \Wjets{}, two shape uncertainties apply to the calculated reweighting corrections, corresponding to uncertainties in the $P_0$ and $P_1$ parameters of the function $f(\ST{}) = P_0 + \exp(P_1 \ST{})$ introduced in Eq.~(\ref{eq:fit_func}).
A uniform 40\% normalization uncertainty is assigned to $t\bar{t}$ events based on the pre-fit uncertainty at high \ST{} for these events.  The limit is not sensitive to the normalization of \Wjets{} so it is included as a free parameter in the final fit.

Uncertainties due to the choice of QCD scales are estimated for the main backgrounds, from \Wjets{}, $t\bar{t}$, and single-top events.
These uncertainties are evaluated with a six-point variation of the chosen renormalization and factorization scales, evaluated by either doubling or halving each, or both, from the nominal values.
The uncertainty due to the choice of PDF is estimated by using alternative PDFs and also the \PDFforLHC[15] combined PDF set~\cite{Butterworth_2016}. For the \Wjets{} samples, the uncertainties associated with matching the matrix elements to the parton showers, and merging different jet multiplicities into an inclusive sample, are estimated from comparisons between the samples with altered CKKW merging/matching scales and QSF parameter values listed in Section~\ref{sec:samples}.

The effects of using an alternative MC generator, ISR and FSR model, parton shower model, and shower matching scheme are estimated by comparing the nominal $t\bar{t}$ sample with alternative samples, as explained in Section~\ref{sec:samples}. Similar uncertainties were evaluated for the single-top events. In addition, the uncertainty due to the choice of scheme to eliminate interference and overlap between $tW$ and \ttbar production is evaluated by comparing the nominal single-top MC sample produced with the DR scheme with the alternative sample produced with the DS scheme~\cite{Frixione:2008yi,ATL-PHYS-PUB-2016-020,ATL-PHYS-PUB-2021-042}.

The analysis also has a small contribution from the single-top and
multĳet backgrounds as well as other backgrounds, which include \Zjets{},
$ttV$, and diboson ($VV$ and $VH$) events. An uncertainty of 10\% is assigned to the cross-section for the ``other backgrounds'' sample. Studies showed that varying the uncertainties for the single-top and multĳet backgrounds
between 10\% and 100\% had a negligible impact on the VLQ mass limit
expected in the absence of a signal, so a conservative uncertainty of
100\% is applied to these backgrounds.

All systematic uncertainties are found to have negligible impact on the sensitivity compared to the statistical uncertainty due to the data sample size.


%
%
\section{Statistical analysis and results}
\label{sec:result}

The statistical analysis tests for the presence of VLQ pair production
by performing a fit over the two SRs. The fit utilizes the reconstructed mass of
the leptonically decaying VLQ, $m_\text{VLQ}^\text{lep}$, as the fit variable of choice due
to the discriminating power provided by the VLQ mass peak not present in SM
backgrounds.

The SRs are fitted to maximize a binned likelihood function
${\cal L}(\mu,\theta)$. This likelihood function ${\cal L}(\mu,\theta)$
is constructed from a product of Poisson probability terms
over all bins considered in the analysis. This function
depends on the parameter of interest, the signal strength
parameter $\mu$, which is a multiplicative factor applied to the
theoretical signal production cross-section, and $\theta$,
a set of nuisance parameters (NPs) which account for the effects of
systematic uncertainties on the number of observed signal and background
events. The NPs are implemented as Gaussian or log-normal priors in
the likelihood. Therefore, the number of events expected in a
particular bin depends on the parameters $\mu$ and $\theta$.
The signal strength $\mu$ is determined by $\mu=\sigma^{\text{test}}/\sigma^{\text{theory}}$,
where $\sigma^{\text{test}}$ is the cross-section for VLQ pair production used in
the fit and $\sigma^{\text{theory}}$ is the theoretical prediction.
The NPs $\theta$ provide variations in the signal and background event counts
consistent with systematic uncertainties. The fitted values of $\theta$
account for the deviations from the nominal expectation that provides
the best fit to data.

To determine the statistical significance of our results, the test
statistic $q_{\mu}$ is taken to be the profile likelihood ratio:
$q_{\mu}=-2\ln{({\cal L}(\mu, \hat{\hat{\theta}}_{\mu})/{\cal L}(\hat{\mu},\hat{\theta}))}$
where $\hat{\mu}$ and $\hat{\theta}$ are the parameter values that
maximize the likelihood function, subject to the condition $0 \leq \hat{\mu} \leq \mu$,
and $\hat{\hat{\theta}}_{\mu}$ are the NP values that maximize the likelihood
function for a given value of $\mu$.

\begin{figure}[tb]
\centering
\subfloat[]{
\includegraphics[width=0.45\textwidth]{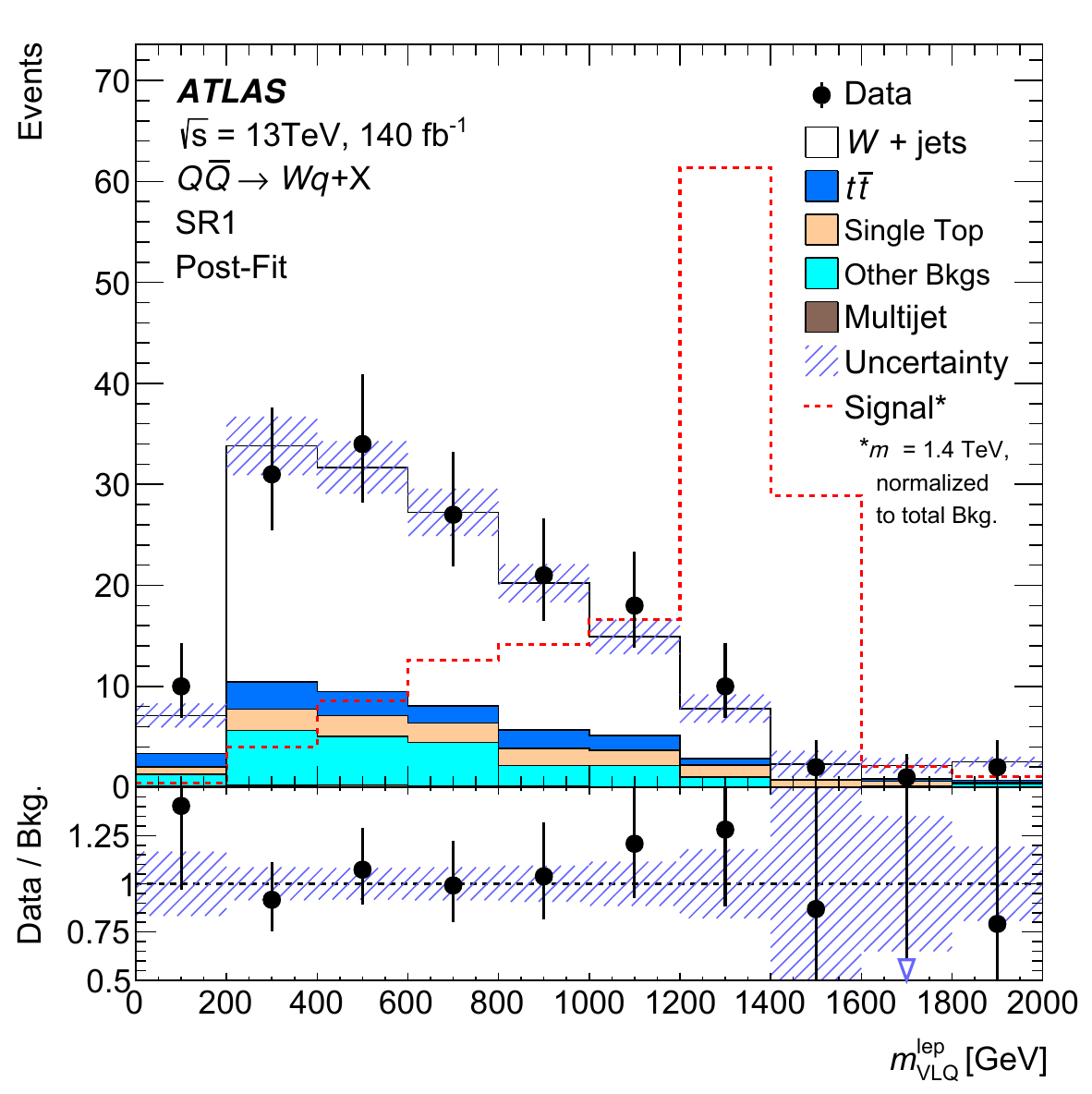}
\label{fig:postFit_SR1}
}
\subfloat[]{
\includegraphics[width=0.45\textwidth]{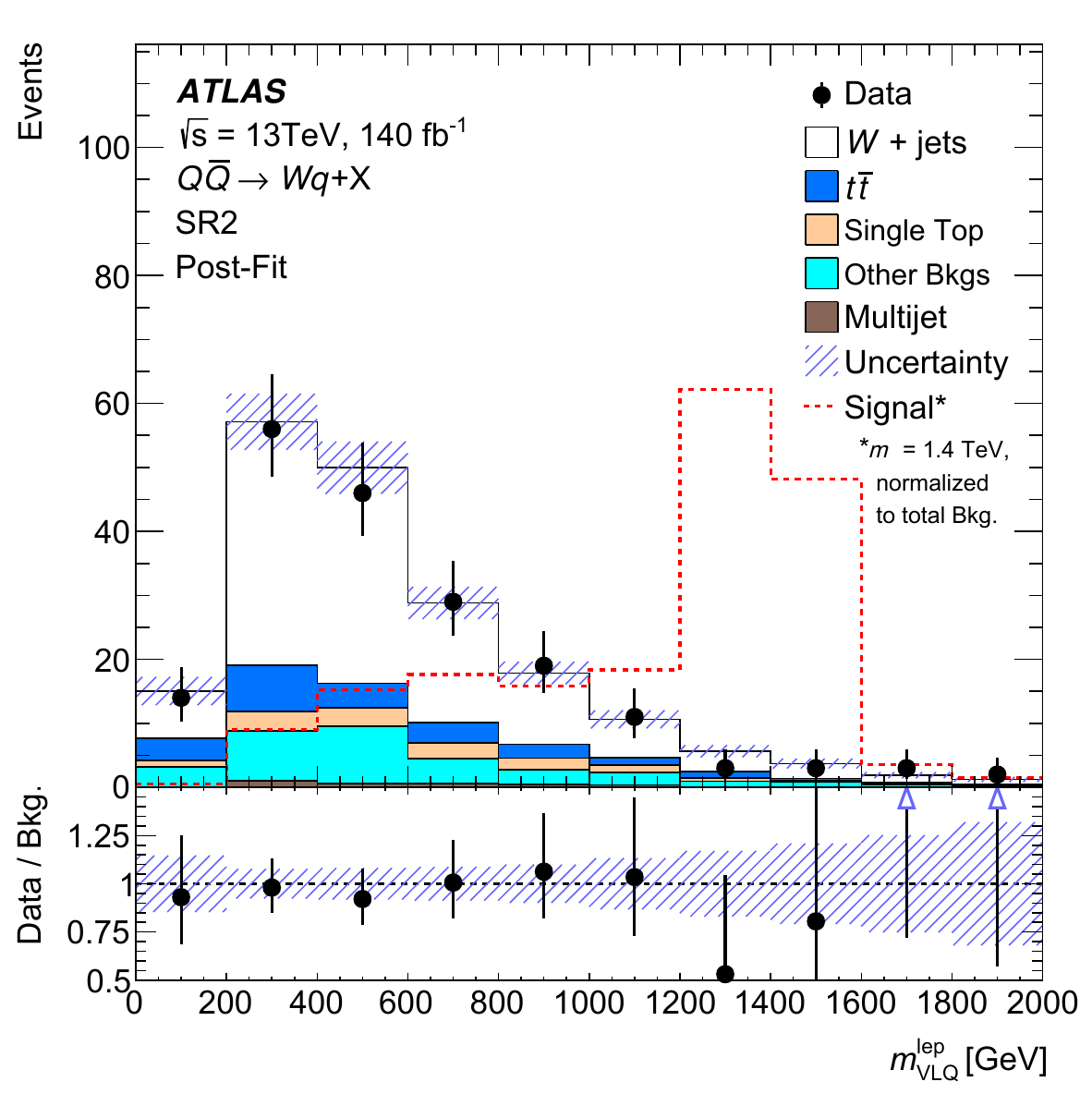}
\label{fig:postFit_SR2}
}
\caption[Post-fit results]{
Distribution for the reconstructed mass of the leptonically decaying VLQ, $m_\text{VLQ}^\text{lep}$, in \protect\subref{fig:postFit_SR1} \SR{1} and \protect\subref{fig:postFit_SR2} \SR{2} after the fit using the background-only hypothesis.
For comparison, the shape of the VLQ signal distribution is overlaid for $\BR{\ensuremath{Q}  \to Wq}=1$ with \mVLQ{}=1400~\GeV{} and normalized to the total background.
The uncertainty bands include all systematic uncertainties described in Section~\ref{sec:systematics}.
An arrow in the bottom panel indicates that the given data point falls outside the range of the plot. }
\label{fig:postFitResults}
\end{figure}

\begin{figure}[tb]
\centering
\subfloat[]{
\includegraphics[width=0.38\pdfpagewidth]{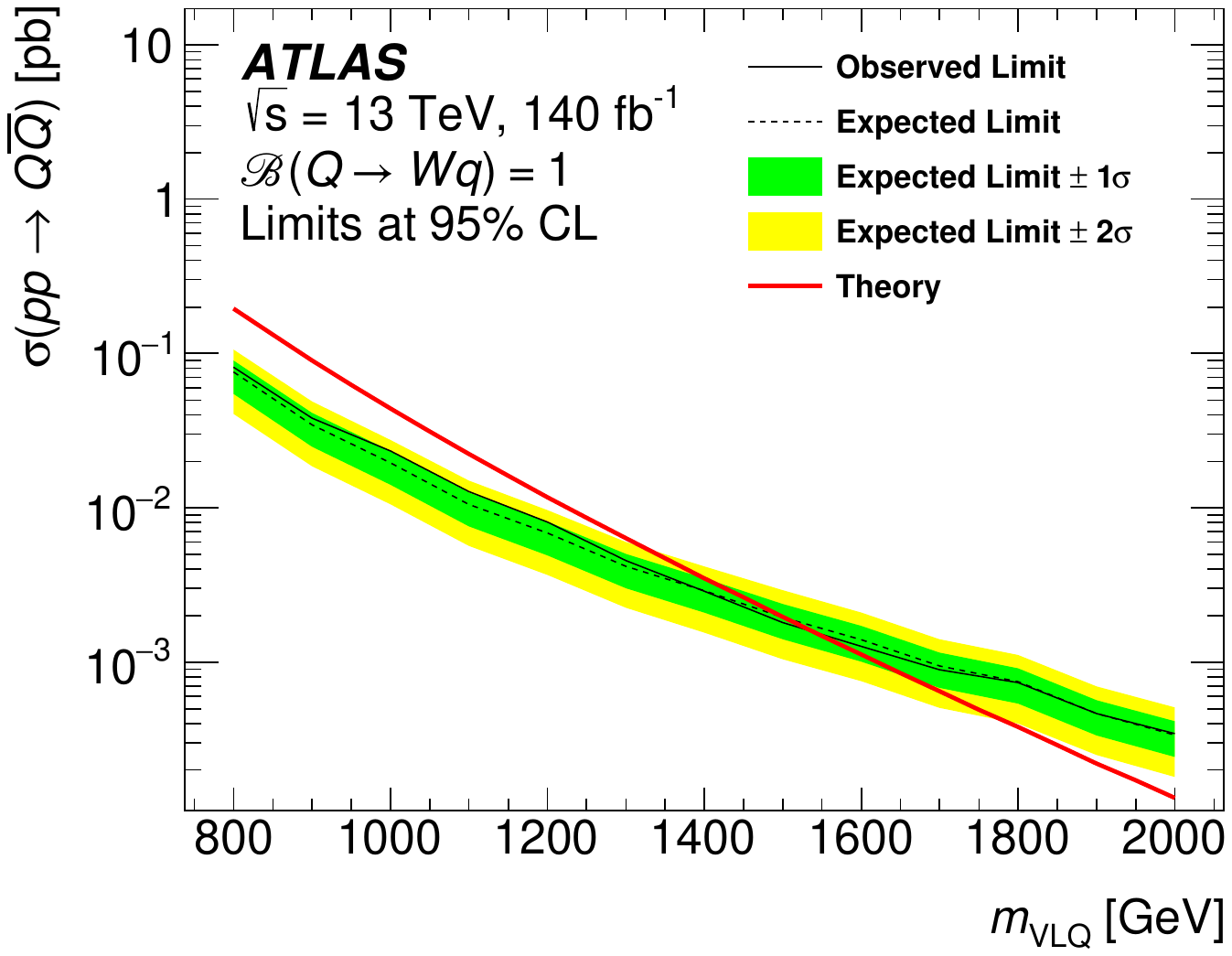}
\label{fig:xsLimit_0_100_0}
}
\subfloat[]{
\includegraphics[width=0.38\pdfpagewidth]{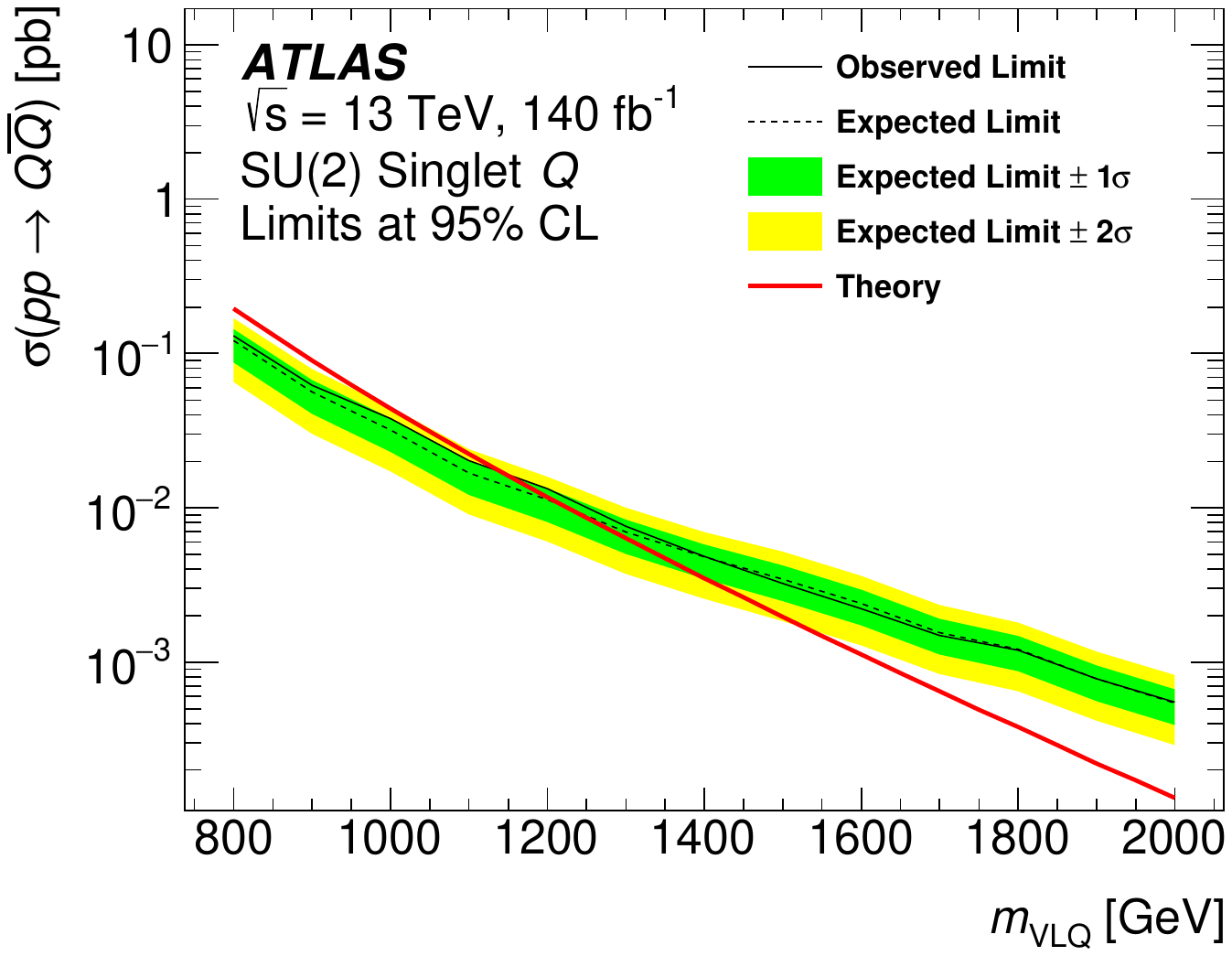}
\label{fig:xsLimit_25_50_25}
}
\caption{Expected (dashed black line) and observed (solid black line) 95\% CL upper limits on the VLQ pair-production cross-section as a function of the VLQ mass for \protect\subref{fig:xsLimit_0_100_0} $\BR{\ensuremath{Q}  \to Wq}=1$ and \protect\subref{fig:xsLimit_25_50_25} the SU(2) singlet model. The green and yellow bands correspond to $\pm$1 and $\pm$2 standard deviations around the expected limit, respectively. The thin red line shows the theoretical prediction for the given model. }
\label{fig:crossSectionLimits}
\end{figure}

The results of the fit in the two \SR{}s for the background-only hypothesis, which assumes $\mu$ = 0, are shown in Figure~\ref{fig:postFitResults}. The corresponding yields are listed in Table~\ref{tab:yields}. From this fit one can see that the data are in good agreement with the SM background prediction. The expected $\QQbar$ event yields for three different mass points in the SRs for the scenario $\BR{Q \to Wq}=1$ are listed in Table~\ref{tab:VLQYields}. To maximize the sensitivity of the search, the two signal regions are combined statistically, allowing the limits on the signal strength to be translated into limits on the total cross-section.
Upper limits at 95\% confidence level (CL) are set on the pair-production cross-section for VLQs with masses from 800~\GeV{} to
2000~\GeV{}. Figure~\ref{fig:crossSectionLimits} shows the results for two benchmark scenarios, one with $\BR{\ensuremath{Q} \to Wq}=1$ and the other with $\mathcal{B}(\ensuremath{Q}\to Wq:Zq:Hq) = 0.5:0.25:0.25$ in the SU(2) singlet model.
Comparing the observed limits on the cross-section with the theory prediction, VLQs with masses below \MassLimitObsWq{} are excluded for $\BR{\ensuremath{Q} \to Wq}=1$ and VLQs with masses below  \MassLimitObsSinglet{} are excluded for the SU(2) singlet model, while the expected limits for these two cases allow $m_\text{VLQ} >$~\MassLimitExpWq{} and \MassLimitExpSinglet{}, respectively. This is an improvement of 840~\GeV{} on the previous ATLAS limit~\cite{EXOT-2014-10} and 685~\GeV{} higher than the latest CMS limit~\cite{CMS-B2G-12-016}.

\begin{table}[tb]
\centering
\caption{Event yields in the two SRs after the fit to the data under the background-only hypothesis. The
uncertainties include statistical and systematic uncertainties.}
\label{tab:yields}
\begin{tabular}{l c c }
\toprule
Process & {SR1} & {SR2}\\
\midrule
\Wjets   & 104\phantom{.0} $\pm$ 20\phantom{.0}    & 124\phantom{.0} $\pm$ 23\phantom{.0} \\
$t\bar{t}$   & \phantom{0}13\phantom{.0} $\pm$ 11\phantom{.0}     & \phantom{0}23\phantom{.0} $\pm$ 15\phantom{.0} \\
Single Top   & \phantom{0}11\phantom{.0} $\pm$ 16\phantom{.0}     & \phantom{0}12\phantom{.0} $\pm$ 15\phantom{.0}  \\
Other Bkgs   & \phantom{0}21.3 $\pm$ \phantom{0}3.0  & \phantom{0}30\phantom{.0} $\pm$ \phantom{0}4\phantom{.0} \\
Multijet     & \phantom{00}0.9 $\pm$ \phantom{0}0.9       & \phantom{00}3.4 $\pm$ \phantom{0}3.3 \\
\midrule
Total        & 150\phantom{.0} $\pm$ 10\phantom{.0}    & 192\phantom{.0} $\pm$ 12\phantom{.0} \\
Data         & 156\phantom{.0 $\pm$ 00.0}            & 186\phantom{.0 $\pm$ 00.0} \\
\bottomrule
\end{tabular}
\end{table}

\begin{table}[tb]
\centering
\caption{Expected $\QQbar$ event yields in the SRs for the scenario $\BR{Q \to Wq}=1$.  The uncertainties include statistical and systematic uncertainties.}
\label{tab:VLQYields}

\begin{tabular}{c c c}
\toprule
$\mVLQ~[\GeV]$  & SR1 & SR2 \\
\midrule
1300 & 31.0 $\pm$ 2.9 & 15.0 $\pm$ 1.2  \\
1400 & 19.3 $\pm$ 2.2 & ~~9.4 $\pm$ 0.8 \\
1500 & 11.9 $\pm$ 1.7 & ~~5.8 $\pm$ 0.5 \\
\bottomrule
\end{tabular}
\end{table}

\begin{figure}[tb]
\centering
\subfloat[]{
\includegraphics[width=0.38\pdfpagewidth]{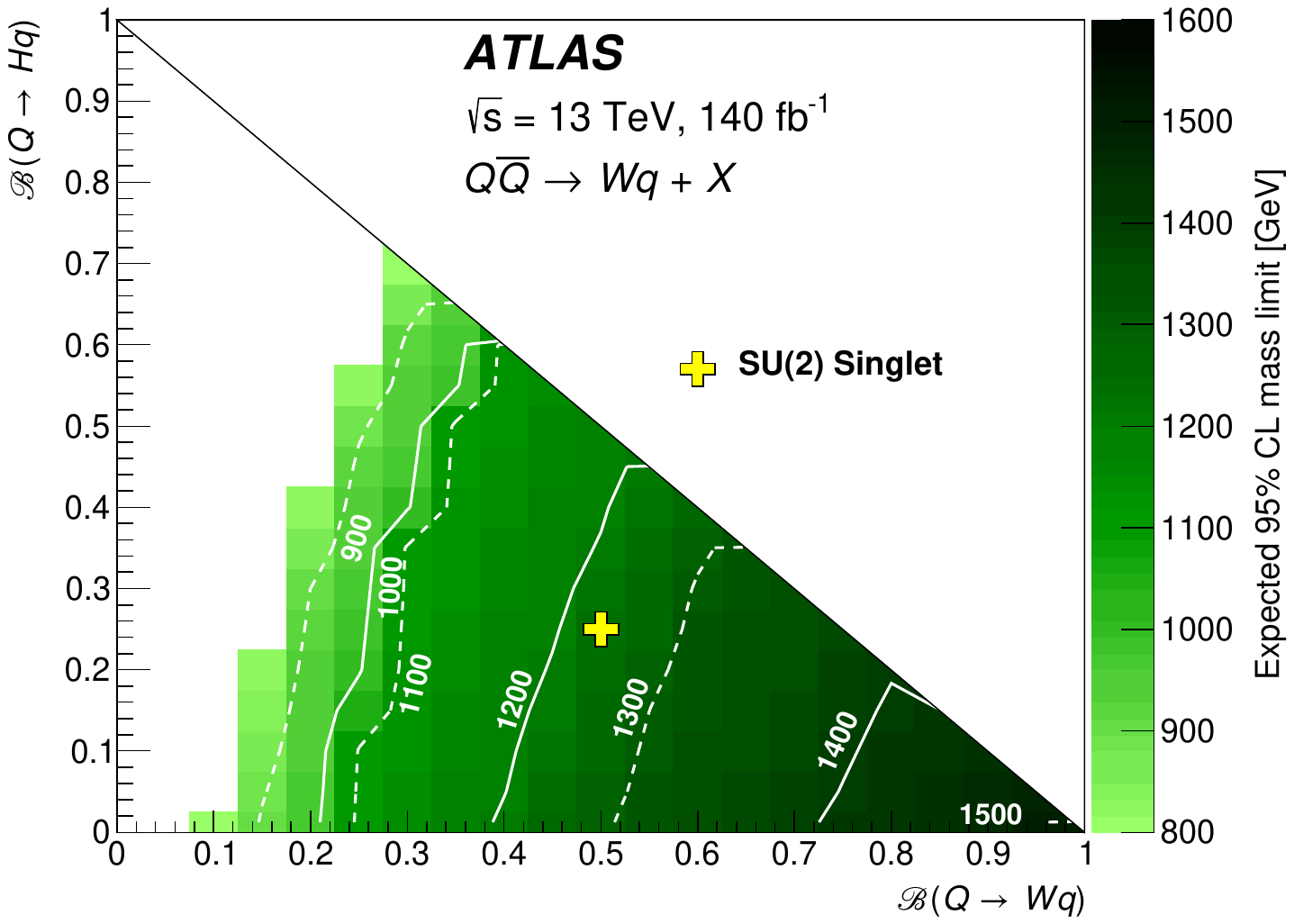}
\label{fig:tempExp}
}
\subfloat[]{
\includegraphics[width=0.38\pdfpagewidth]{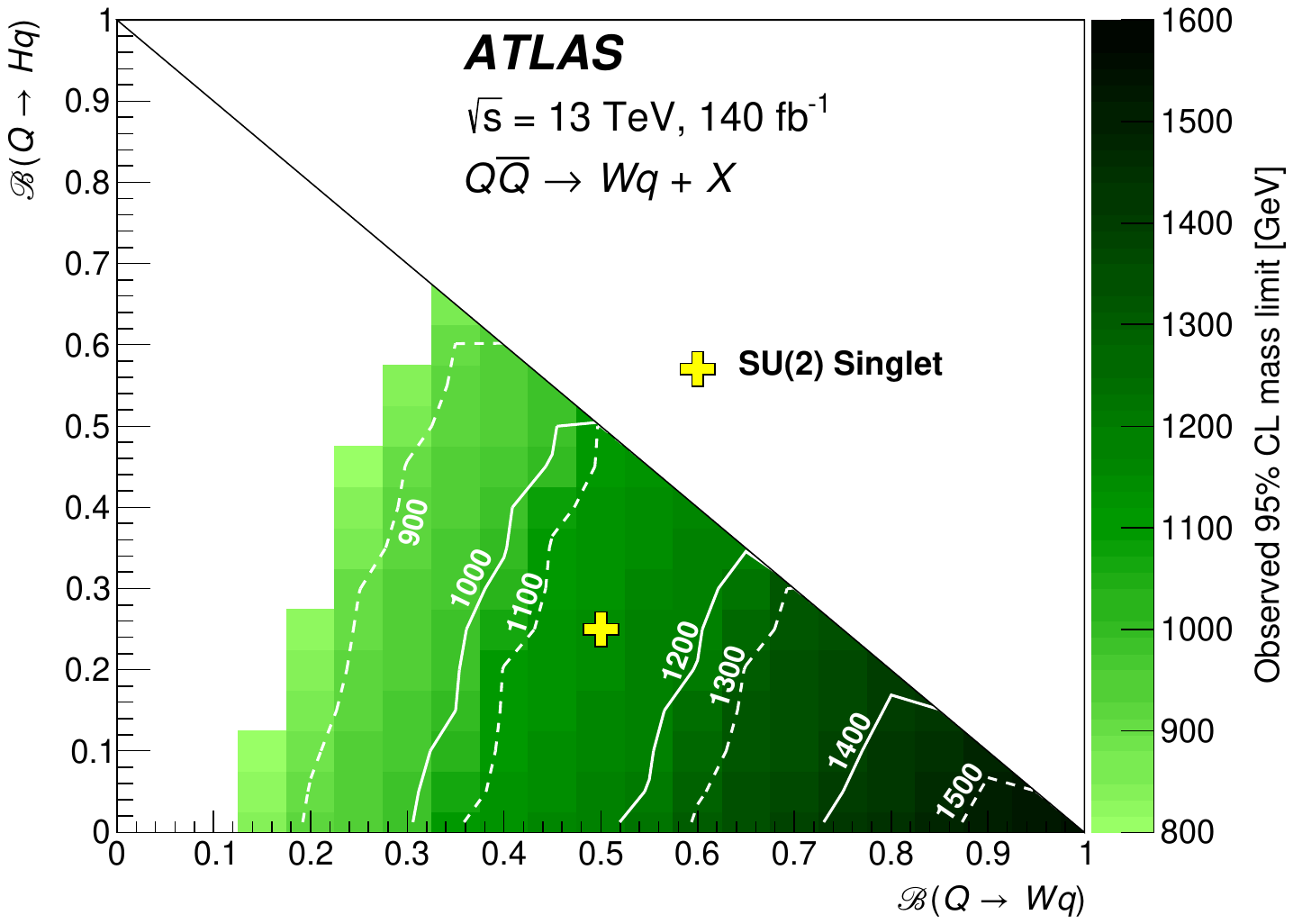}
\label{fig:tempObs}
}
\caption{\protect\subref{fig:tempExp} Expected and \protect\subref{fig:tempObs} observed lower limit on the VLQ mass for various branching ratio ($\mathcal{B}$) configurations at 95\% CL. The $x$-axis shows the branching ratio for decay to a $W$ boson, while the $y$-axis shows the branching ratio for decay to a Higgs boson. The branching ratio to a $Z$ boson is determined by the requirement that $\mathcal{B}(Q \to Zq) = 1 - \mathcal{B}(Q \to Wq) - \mathcal{B}(Q \to Hq)$. The white region above the solid black line is excluded due to the sum of the individual branching ratios being greater than one.  In the white region below this line the lower limit on the VLQ mass is below 800~\GeV. The yellow marker indicates the branching ratios for the SU(2) singlet scenario.}
\label{fig:BRTemp}
\end{figure}
Although this analysis was optimized to search for VLQs that decay into a $W$ boson and a light quark, it also has some sensitivity to signal events with neutral-current decays via a Higgs or $Z$ boson.  The sensitivity to $Z$ boson decays is larger given the similarities in the topology of a hadronically decaying $Z$ and $W$ boson. The likelihood of this occurring for a Higgs boson is much lower given the parameters used in the tagging algorithm discussed in Section \ref{sec:object}. Assuming no other decay modes, the signal MC samples are reweighted to obtain different branching ratio combinations, which can then be used in the final fit. Figure~\ref{fig:BRTemp} shows 95\% CL
lower limits on the mass of the VLQ for various branching ratios.

\FloatBarrier


%

%

%
\section{Conclusion}
\label{sec:conclusion}

A search for the pair production of heavy vectorlike quarks ($\QQbar{}$), where each subsequently decays  into a $W$ boson and a light quark, was performed using \lumi{} of 13~\TeV $pp$ collision data recorded with the ATLAS detector during \RunTwo of the LHC.
The analysis specifically targets events in the semileptonic final state, requiring the reconstruction of one leptonically decaying $W$ boson and one hadronically decaying $W$ boson.
The final state is characterized by a high-transverse-momentum isolated electron or muon, substantial missing transverse momentum, multiple jets, and a large-radius jet identified as originating from the hadronic decay of a $W$ boson.
No significant excess is discovered in the reconstructed vectorlike quark mass distribution, and 95\% CL upper limits are derived on the signal production cross section as a function of the vectorlike quark mass.
The analysis sets constraints on the mass of the vectorlike quarks for various combinations of their decays to SM bosons and light SM quarks.
Notably, masses below \MassLimitObsWq{} are excluded at 95\% CL for $\mathcal{B}(\ensuremath{Q}\to Wq) = 1$, and masses below \MassLimitObsSinglet{} are excluded for the scenario with $\mathcal{B}(\ensuremath{Q}\to Wq:Zq:Hq) = 0.5:0.25:0.25$.
With increased center-of-mass energy and integrated luminosity, as well as improved analysis tools, the exclusion limit for $\mathcal{B}(\ensuremath{Q}\to Wq) = 1$ is raised by 840~\GeV and 685~\GeV compared to the \RunOne analyses conducted by ATLAS and CMS, respectively.


%

%
\section*{Acknowledgments}
%

%
%

%
%

We thank CERN for the very successful operation of the LHC and its injectors, as well as the support staff at
CERN and at our institutions worldwide without whom ATLAS could not be operated efficiently.

The crucial computing support from all WLCG partners is acknowledged gratefully, in particular from CERN, the ATLAS Tier-1 facilities at TRIUMF/SFU (Canada), NDGF (Denmark, Norway, Sweden), CC-IN2P3 (France), KIT/GridKA (Germany), INFN-CNAF (Italy), NL-T1 (Netherlands), PIC (Spain), RAL (UK) and BNL (USA), the Tier-2 facilities worldwide and large non-WLCG resource providers. Major contributors of computing resources are listed in Ref.~\cite{ATL-SOFT-PUB-2023-001}.

We gratefully acknowledge the support of ANPCyT, Argentina; YerPhI, Armenia; ARC, Australia; BMWFW and FWF, Austria; ANAS, Azerbaijan; CNPq and FAPESP, Brazil; NSERC, NRC and CFI, Canada; CERN; ANID, Chile; CAS, MOST and NSFC, China; Minciencias, Colombia; MEYS CR, Czech Republic; DNRF and DNSRC, Denmark; IN2P3-CNRS and CEA-DRF/IRFU, France; SRNSFG, Georgia; BMBF, HGF and MPG, Germany; GSRI, Greece; RGC and Hong Kong SAR, China; ISF and Benoziyo Center, Israel; INFN, Italy; MEXT and JSPS, Japan; CNRST, Morocco; NWO, Netherlands; RCN, Norway; MNiSW, Poland; FCT, Portugal; MNE/IFA, Romania; MESTD, Serbia; MSSR, Slovakia; ARIS and MVZI, Slovenia; DSI/NRF, South Africa; MICIU/AEI, Spain; SRC and Wallenberg Foundation, Sweden; SERI, SNSF and Cantons of Bern and Geneva, Switzerland; NSTC, Taipei; TENMAK, T\"urkiye; STFC/UKRI, United Kingdom; DOE and NSF, United States of America.

Individual groups and members have received support from BCKDF, CANARIE, CRC and DRAC, Canada; CERN-CZ, FORTE and PRIMUS, Czech Republic; COST, ERC, ERDF, Horizon 2020, ICSC-NextGenerationEU and Marie Sk{\l}odowska-Curie Actions, European Union; Investissements d'Avenir Labex, Investissements d'Avenir Idex and ANR, France; DFG and AvH Foundation, Germany; Herakleitos, Thales and Aristeia programmes co-financed by EU-ESF and the Greek NSRF, Greece; BSF-NSF and MINERVA, Israel; Norwegian Financial Mechanism 2014-2021, Norway; NCN and NAWA, Poland; La Caixa Banking Foundation, CERCA Programme Generalitat de Catalunya and PROMETEO and GenT Programmes Generalitat Valenciana, Spain; G\"{o}ran Gustafssons Stiftelse, Sweden; The Royal Society and Leverhulme Trust, United Kingdom.

In addition, individual members wish to acknowledge support from CERN: European Organization for Nuclear Research (CERN PJAS); Chile: Agencia Nacional de Investigaci\'on y Desarrollo (FONDECYT 1190886, FONDECYT 1230812, FONDECYT 1230987); China: Chinese Ministry of Science and Technology (MOST-2023YFA1605700), National Natural Science Foundation of China (NSFC - 12175119, NSFC 12275265, NSFC-12075060); Czech Republic: Czech Science Foundation (GACR - 24-11373S), Ministry of Education Youth and Sports (FORTE CZ.02.01.01/00/22\_008/0004632), PRIMUS Research Programme (PRIMUS/21/SCI/017); EU: H2020 European Research Council (ERC - 101002463); European Union: European Research Council (ERC - 948254, ERC 101089007), Horizon 2020 Framework Programme (MUCCA - CHIST-ERA-19-XAI-00), European Union, Future Artificial Intelligence Research (FAIR-NextGenerationEU PE00000013), Italian Center for High Performance Computing, Big Data and Quantum Computing (ICSC, NextGenerationEU); France: Agence Nationale de la Recherche (ANR-20-CE31-0013, ANR-21-CE31-0013, ANR-21-CE31-0022, ANR-22-EDIR-0002), Investissements d'Avenir Labex (ANR-11-LABX-0012); Germany: Baden-Württemberg Stiftung (BW Stiftung-Postdoc Eliteprogramme), Deutsche Forschungsgemeinschaft (DFG - 469666862, DFG - CR 312/5-2); Italy: Istituto Nazionale di Fisica Nucleare (ICSC, NextGenerationEU), Ministero dell'Università e della Ricerca (PRIN - 20223N7F8K - PNRR M4.C2.1.1); Japan: Japan Society for the Promotion of Science (JSPS KAKENHI JP21H05085, JSPS KAKENHI JP22H01227, JSPS KAKENHI JP22H04944, JSPS KAKENHI JP22KK0227); Netherlands: Netherlands Organisation for Scientific Research (NWO Veni 2020 - VI.Veni.202.179); Norway: Research Council of Norway (RCN-314472); Poland: Ministry of Science and Higher Education (IDUB AGH, POB8, D4 no 9722), Polish National Agency for Academic Exchange (PPN/PPO/2020/1/00002/U/00001), Polish National Science Centre (NCN 2021/42/E/ST2/00350, NCN OPUS nr 2022/47/B/ST2/03059, NCN UMO-2019/34/E/ST2/00393, UMO-2020/37/B/ST2/01043, UMO-2021/40/C/ST2/00187, UMO-2022/47/O/ST2/00148, UMO-2023/49/B/ST2/04085); Slovenia: Slovenian Research Agency (ARIS grant J1-3010); Spain: Generalitat Valenciana (Artemisa, FEDER, IDIFEDER/2018/048), Ministry of Science and Innovation (MCIN \& NextGenEU PCI2022-135018-2, MICIN \& FEDER PID2021-125273NB, RYC2019-028510-I, RYC2020-030254-I, RYC2021-031273-I, RYC2022-038164-I), PROMETEO and GenT Programmes Generalitat Valenciana (CIDEGENT/2019/023, CIDEGENT/2019/027); Sweden: Swedish Research Council (Swedish Research Council 2023-04654, VR 2018-00482, VR 2022-03845, VR 2022-04683, VR 2023-03403, VR grant 2021-03651), Knut and Alice Wallenberg Foundation (KAW 2018.0157, KAW 2018.0458, KAW 2019.0447, KAW 2022.0358); Switzerland: Swiss National Science Foundation (SNSF - PCEFP2\_194658); United Kingdom: Leverhulme Trust (Leverhulme Trust RPG-2020-004), Royal Society (NIF-R1-231091); United States of America: U.S. Department of Energy (ECA DE-AC02-76SF00515), Neubauer Family Foundation.

%
%


%
%
%
%
%
%
%

\clearpage
\printbibliography
\clearpage
 
\begin{flushleft}
\hypersetup{urlcolor=black}
{\Large The ATLAS Collaboration}

\bigskip

\AtlasOrcid[0000-0002-6665-4934]{G.~Aad}$^\textrm{\scriptsize 104}$,
\AtlasOrcid[0000-0001-7616-1554]{E.~Aakvaag}$^\textrm{\scriptsize 17}$,
\AtlasOrcid[0000-0002-5888-2734]{B.~Abbott}$^\textrm{\scriptsize 123}$,
\AtlasOrcid[0000-0002-0287-5869]{S.~Abdelhameed}$^\textrm{\scriptsize 119a}$,
\AtlasOrcid[0000-0002-1002-1652]{K.~Abeling}$^\textrm{\scriptsize 56}$,
\AtlasOrcid[0000-0001-5763-2760]{N.J.~Abicht}$^\textrm{\scriptsize 50}$,
\AtlasOrcid[0000-0002-8496-9294]{S.H.~Abidi}$^\textrm{\scriptsize 30}$,
\AtlasOrcid[0009-0003-6578-220X]{M.~Aboelela}$^\textrm{\scriptsize 45}$,
\AtlasOrcid[0000-0002-9987-2292]{A.~Aboulhorma}$^\textrm{\scriptsize 36e}$,
\AtlasOrcid[0000-0001-5329-6640]{H.~Abramowicz}$^\textrm{\scriptsize 154}$
% ...
\bigskip
\\

% ...
$^{17}$Department for Physics and Technology, University of Bergen, Bergen; Norway.\\
% ...
$^{30}$Physics Department, Brookhaven National Laboratory, Upton NY; United States of America.\\
% ...
$^{36}$$^{(a)}$Facult\'e des Sciences Ain Chock, Universit\'e Hassan II de Casablanca;$^{(b)}$Facult\'{e} des Sciences, Universit\'{e} Ibn-Tofail, K\'{e}nitra;$^{(c)}$Facult\'e des Sciences Semlalia, Universit\'e Cadi Ayyad, LPHEA-Marrakech;$^{(d)}$LPMR, Facult\'e des Sciences, Universit\'e Mohamed Premier, Oujda;$^{(e)}$Facult\'e des sciences, Universit\'e Mohammed V, Rabat;$^{(f)}$Institute of Applied Physics, Mohammed VI Polytechnic University, Ben Guerir; Morocco.\\
% ...
$^{45}$Physics Department, Southern Methodist University, Dallas TX; United States of America.\\
% ...
$^{50}$Fakult\"{a}t Physik , Technische Universit{\"a}t Dortmund, Dortmund; Germany.\\
% ...
$^{56}$II. Physikalisches Institut, Georg-August-Universit\"{a}t G\"ottingen, G\"ottingen; Germany.\\
% ...
$^{104}$CPPM, Aix-Marseille Universit\'e, CNRS/IN2P3, Marseille; France.\\
% ...
$^{119}$$^{(a)}$New York University Abu Dhabi, Abu Dhabi;$^{(b)}$United Arab Emirates University, Al Ain; United Arab Emirates.\\
% ...
$^{123}$Homer L. Dodge Department of Physics and Astronomy, University of Oklahoma, Norman OK; United States of America.\\
% ...
$^{154}$Raymond and Beverly Sackler School of Physics and Astronomy, Tel Aviv University, Tel Aviv; Israel.\\
% ...
$^{a}$ Also Affiliated with an institute covered by a cooperation agreement with CERN.\\
$^{b}$ Also at An-Najah National University, Nablus; Palestine.\\
$^{c}$ Also at Borough of Manhattan Community College, City University of New York, New York NY; United States of America.\\
$^{d}$ Also at Center for Interdisciplinary Research and Innovation (CIRI-AUTH), Thessaloniki; Greece.\\
$^{e}$ Also at Centro Studi e Ricerche Enrico Fermi; Italy.\\
$^{f}$ Also at CERN, Geneva; Switzerland.\\
% ...

\end{flushleft}


%

%
%

%

%
%
%
%

\end{document}